\documentclass[%
preprint,
superscriptaddress,
 amsmath,amssymb,
 aps,
prb,
]{revtex4-2}

\usepackage[english]{babel}
\usepackage{graphicx}
\usepackage{dcolumn}
\usepackage{bm}
\usepackage{hyperref}
\usepackage{enumitem}

\hypersetup{pdfstartview={FitH},pdfpagemode={UseNone}, breaklinks=true,colorlinks=true,allcolors=blue,bookmarksopen=true, pdfnewwindow=true}
\usepackage[all]{hypcap}
\usepackage{physics}
\usepackage{cleveref}
\crefname{figure}{Fig.}{Figs.}
\usepackage{bbold}

\newcommand{\doubleitem}{%
  \begingroup
  \stepcounter{enumi}%
  \edef\tmp{\theenumi, }%
  \stepcounter{enumi}
  \edef\tmp{\endgroup\noexpand\item[\tmp\labelenumi]}%
  \tmp}
  
\crefname{enumi}{objective}{objectives}
\Crefname{enumi}{Objective}{Objectives}

\newcommand{\yb}{Yb\textsuperscript{3+}}
\newcommand{\er}{Er\textsuperscript{3+}}

\newcommand{\supbasis}{Supp.~S1}
\newcommand{\supsymvec}{Supp.~S2}
\newcommand{\suporigin}{Supp.~S3}
\newcommand{\supamb}{Supp.~S4}

\begin{document}
\sloppy


\title{Optimization of metasurfaces for lasing with symmetry constraints on the modes}

\author{Matthew Parry}
\email{Matthew.Parry@anu.edu.au}
\affiliation{ARC Centre of Excellence for Transformative Meta-Optical Systems (TMOS), Department of Electronic Materials Engineering,  Research School of Physics, Australian National University, Canberra, ACT 2601, Australia}%

\author{Kenneth B. Crozier}
\affiliation{ARC Centre of Excellence for Transformative Meta-Optical Systems (TMOS), Department of Electrical and Electronic Engineering, School of Physics, University of Melbourne, Melbourne, Victoria 3010, Australia}
 
\author{Andrey A. Sukhorukov}%


\author{Dragomir N. Neshev}
\affiliation{ARC Centre of Excellence for Transformative Meta-Optical Systems (TMOS), Department of Electronic Materials Engineering,  Research School of Physics, Australian National University, Canberra, ACT 2601, Australia}%


\date{\today}

\begin{abstract}
The development of active metasurface systems, such as lasing metasurfaces, requires the optimization of multiple modes at the absorption and lasing wavelength bands, including their quality factor, mode profile and angular dispersion. Often, these requirements are contradictory and impossible to obtain with conventional design techniques. 
Importantly, the properties of the eigenmodes of a metasurface are directly linked to their symmetry, which offers an opportunity to explore mode symmetry as an objective in optimization routines for active metasurface design. Here, we propose and numerically demonstrate a novel multi-objective optimization technique based on symmetry projection operators to quantify the symmetry of the metasurface eigenmodes. We present, as an example, the optimization of a lasing metasurface based on up-converting nano-particles. Our technique allows us to optimize the absorption mode dispersion, as well as the directionality of the lasing emission and therefore offers advantages for novel lasing systems with high directionality and low lasing threshold.
\end{abstract}

\maketitle



\section{Introduction}

Active metasurfaces (MSs) have received a great deal of interest over recent years and in particular with applications such as micro and nano-lasers.~\cite{Zhou:2013-506:NNANO,Yang:2015-6939:NCOM,Wang:2017-889:NNANO,Wang:2018-4549:NANL,Fernandez-Bravo:2019-1172:NMAT} Various MS lasers have been demonstrated, including vortex lasers~\cite{Seghilani:2016-38156:SRP,
Sroor:2020-498:NPHOT,Wen2021,Sroor2020,MermetLyaudoz2023} and micro-lasers with strong directional light-emission.~\cite{MermetLyaudoz2023} Most MS lasers have utilized optical pumping, where the MS properties have to be optimized simultaneously for both the pump and emission wavelengths. However, techniques for rigorous design have remained largely unexplored.

Recently MS lasers have been demonstrated using Up-Converting Nano-Particles (UCNPs)~\cite{Fernandez-Bravo:2019-1172:NMAT}, which contain a matrix of Lanthanides with appropriate sequential two-photon transitions~\cite{Chen:2014-5161:CHR}. In contrast to conventional MS lasers, the absorption band of the UCNPs is at longer wavelengths than the lasing spectral band (\cref{fig:ms}a). This active metasurface system offers unique design complexity as the lasing band is covered with higher-order modes. The work of Ref.~\cite{Fernandez-Bravo:2019-1172:NMAT} achieved lasing at the desired transition using a plasmonic MS. However, the plasmonic MS was only resonant at the lasing wavelength, having no enhancement for the absorption. Furthermore, the mode volume in the gain medial for plasmonic MSs is relatively low, therefore limiting the lasing threshold.   

Dielectric MSs were therefore explored as a possible alternative. A recent work has studied Si dielectric MSs~\cite{Gong2019} covered with Tm UCNPs. The close proximity of the pump and lasing wavelengths in such Tm:UCNPs allowed for a conventional design approach where the electric and magnetic dipole Mie resonances were used to enhance both the absorption and emission bands. 
Alternatively, another recent work used TiO\textsubscript{2} for the resonators~\cite{feng2023dualband} and enhanced two different but closely spaced emission transitions of Yb/Er UCNPs at 660~nm and 540~nm. It is also possible to introduce more complicated engineering of the up-conversion emission, such as the Rashba effect, with MSs~\cite{Tripathi:2023-2228:NANL}.  However, there is still an outstanding problem of how to design MSs that enhance both the absorption and atomic transitions of UCNP-system with distinct and widely spaced transition wavelengths.

Symmetry has recently become an important element of such complex MS design, finding uses in the engineering of resonant leaky modes (such as quasi-bound states in the continuum)~\cite{Huang2023,Koshelev2018}, chirality~\cite{Khaliq2023}, geometric phase~\cite{BalthasarMueller2017,Yu2011} and the effective permittivity and permeability of MSs~\cite{Reinke:2011-66603:PRE}.  The properties of the individual modes themselves can also be determined by symmetry, such as in the engineering of Dirac cones from the overlap of modes of certain symmetries~\cite{Sakoda2012}.  At the same time, inverse design methods are becoming increasingly important for the engineering of specific responses~\cite{Campbell2019,Elsawy2020}.  It is clear then that an important advance can be made in the development of metasurfaces if one is able to introduce symmetry constraints on both the MS design and the selection of modes in inverse design algorithms.  While one can specify the symmetry of a MS in an optimization routine, in order to control the symmetry of the modes one requires a method to quantify the symmetry of vector fields obtained from simulations.  

We, therefore, require new dielectric MS designs that increase the absorption, provide a high-quality factor at the lasing band and optimize the overlap between the two modes. However, this multi-objective optimization is not easy to achieve with conventional routines and requires novel approaches. As noted above, many properties of MSs are determined by the symmetry of the structure~\cite{Padilla2007,Reinke:2011-66603:PRE} and the symmetry of the eigenmodes also play a critical role in determining their dispersion, such as in the formation of Dirac cones~\cite{Sakoda2012,Mei:2012-35141:PRB}.

In this work, we propose and numerically demonstrate a novel approach that is based on projection operators (also known as projectors) to apply symmetry constraints to the excited modes at the different wavelengths and derive efficient designs for MS lasers. We demonstrate this technique for the example of ...

\ldots UCNP based MS lasers, where our technique allows us to select modes with a particular, symmetry dependent, dispersion by quantifying numerically the symmetry of the eigenmodes.   We also use the technique to avoid unwanted MS properties by rejecting overlapping eigenmodes with undesirable symmetry combinations.

\section{Design}\label{sec:design}

\begin{figure}[b]
    \centering
    \includegraphics[width=0.75\columnwidth]{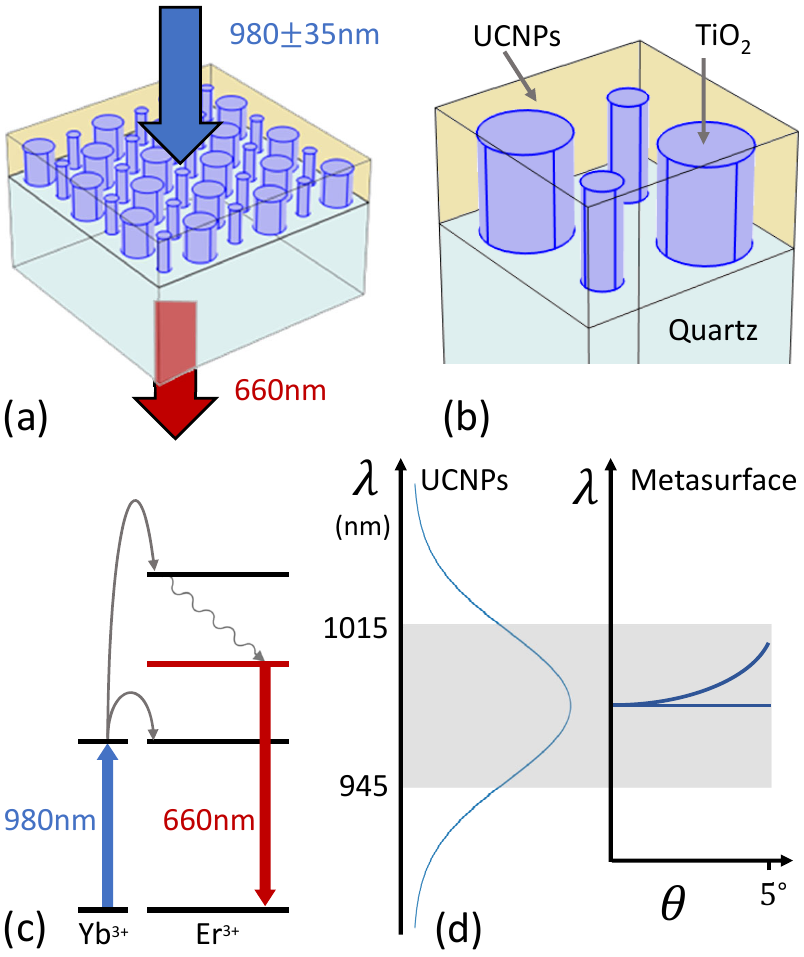}
	\caption[The MS studied]{(a) Conceptual diagram of lasing up-conversion. (b) The design of the MS.  (c) Abbreviated energy level diagram for Yb and Er, showing energy donated by Yb to Er in gray lines and phonons with a wavy line.  (d) (left) Gaussian model of the Yb atom transition at \(980\pm35\)nm, (right) acceptable dispersions for the MS eigenmodes at \(980\pm35\)nm.  The gray region shows the FWHM.}
    \label{fig:ms}
\end{figure}

Extending the work of ref.~\cite{Fernandez-Bravo:2019-1172:NMAT} we designed a MS laser that absorbs at 980nm and emits at 660nm (\cref{fig:ms}a). The active medium for the UCNP MS laser is a layer of nano-particles consisting of a matrix of \yb and \er atoms that infiltrate the TiO\textsubscript{2} MS (\cref{fig:ms}b).  The \yb atoms have a larger cross-section than the \er atoms as well as a single transition at 980nm, which corresponds to two transitions in the \er atoms (\cref{fig:ms}c).  The matrix of Yb and Er atoms in each nanoparticle facilitates the excitation of the Er atom by Yb atoms.  Specifically, two \yb atoms can excite the \er atoms to a higher level, which, following multi-phonon relaxation, can relax to the ground state via 660nm emission~\cite{Chen:2014-5161:CHR}.  For UCNPs alone, the Full Width Half Max (FWHM) of the absorption band is typically 10-20nm, but with dye sensitization, this can be broadened~\cite{Garfield2018, Shao2016, Chen2015}.  In this work, we take the absorption transition to have a bandwidth of \(980\pm35\)nm.  It has been demonstrated~\cite{Fernandez-Bravo:2019-1172:NMAT} that one can cause the 660nm transition to be preferred over the others that are possible by infiltrating a plasmonic MS that has a resonance at 660nm.  In this previous work, however, there was no MS resonance at the wavelength of the \yb transition, and so the absorption was not optimized.  Furthermore, plasmonic resonances have a low field volume, and hence, only UCNPs at the surface of the resonators would have been activated.

The MS design, therefore, requires the optimization of multiple objectives at both the lasing and absorption bands. 
At the lasing band we require:
\vspace{-2mm}
\begin{enumerate}[leftmargin=*]
\itemsep0em
    \item\label{it:ldal}
Minimization of the detuning between the lasing wavelength and the wavelength of the highest quality factor mode, since the highest Q MS mode will dominate lasing;
    \item\label{it:lasQ} That the quality factor of the mode reduces away from the \(\Gamma\)-point, to ensure normal emission;
    \item\label{it:lasfield} Maximization of the ratio of the mode's intensity within the UCNPs to that of the whole unit cell, since only the electromagnetic energy within the UCNPs will enhance the lasing of the UCNPs.
\end{enumerate}
At the absorption band we require:
\begin{enumerate}[resume, leftmargin=*]
\vspace{-2mm}
\parskip0in
\itemsep0em
    \item\label{it:absfield} Maximization of the ratio of the absorption mode's intensity within the UCNPs to the intensity in the whole unit cell.
    \item\label{it:abs} Maximization of the power absorbed within the UCNP region.
    \item\label{it:absangle} Low angular dispersion for the absorption mode so that the acceptance angle for the absorption band is at least as large as the numerical aperture of the optical system (e.g. $\pm 5^\circ$, see \cref{fig:ms}d).
\end{enumerate}

For \cref{it:ldal} and \crefrange{it:lasfield}{it:abs}, we formulated figures of merit (See~\cref{sec:sim}) and multiplied them together to obtain the final figure of merit.  For \cref{it:lasQ} the most common cause of high Q modes not reducing in quality factor away from the \(\Gamma\)-point is the overlap of modes of certain symmetries (See~\cref{sec:sim}) and so if we detect such a mode overlap via the technique outlined in this paper then a FOM of zero is returned.

\begin{figure}[b]
    \centering
    \includegraphics[width=0.75\columnwidth]{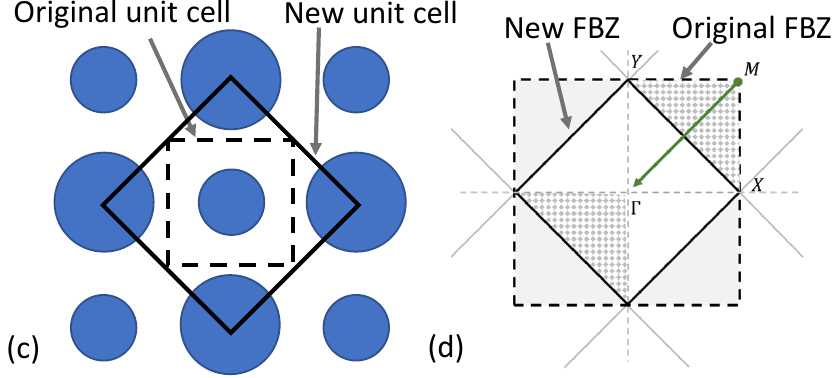}
	\caption[The MS studied]{(a) The unit cell changes as the radii of the cylinders are made non-uniform.  (b) The \(M\)-point of the First Brillouin Zone (FBZ) of the original (uniform radii) MS is folded to the \(\Gamma\)-point of the FBZ of the modified (non-uniform radii) MS.}
    \label{fig:brill}
\end{figure}

\Crefrange{it:lasfield}{it:abs} require a large proportion of the mode fields to be outside of the resonators, which can be achieved by targeting lattice modes.  We, therefore, use a band folding design for the MS, as depicted in \cref{fig:brill}a, as this design increases the number of lattice modes at the \(\Gamma\)-point~\cite{Overvig:2020-35434:PRB}.  By alternating the radii of the resonators in a staggered pattern (\cref{fig:brill}a), modes are band folded from the \(M\)-points to the center of the first Brillouin zone (\cref{fig:brill}b). Our design uses TiO\textsubscript{2} for the resonators on a quartz substrate with a 23~nm thick Indium Tin Oxide layer to enhance adhesion, which are all materials commonly used for MS applications at visible wavelengths.~\cite{Chen:2019-355:NCOM,Sisler:2020-56107:APLP} The height of the UCNP layer is set to be 100~nm higher than the resonators (since one occasionally encounters a mode with a significant proportion of the field just above the resonators), and all other design parameters are set by the optimization routine.

For \cref{it:absangle}, it is possible to create modes with perfectly flat dispersion with Leib and Kagome~\cite{Tang2020,Leykam2018,Leykam2018b} lattices, but these modes exist mostly in the resonators and hence contradict \crefrange{it:lasfield}{it:abs}.  These modes are also dominated by nearest-neighbor interactions and, therefore, will not have the coherence across the lattice required for lasing.  It is also possible to create relatively flat dispersion by tuning the parameters of the MS so that the interaction between multiple modes will give a flat mode.  Optimizing for this, however, would require multiple simulations to check the dispersion of the modes and is thus not practical because of time considerations.

We have therefore opted for a compromise solution of only considering modes at the absorption band of a certain symmetry, known as \(C_{4v}\ E\) symmetry (as explained in~\cref{sec:theory}).  These modes are doubly degenerate and polarization-independent at the \(\Gamma\)-point.  The P-polarized mode has low dispersion along the \(\Gamma-X\) direction and both S and P polarisations have relatively low dispersion along the \(\Gamma-M\) direction.  Although there is dispersion present, this dispersion only has to be less than the bandwidth of the absorptive transition in the UCNPs (\cref{fig:ms}d), and for P-polarized light along the \(\Gamma-X\) direction, this can create an acceptance angle of up to \(15^\circ\).

To target only modes of a particular symmetry requires a method to quantify the symmetry of the modes returned by an eigenfrequency study.  In the following section, we outline our technique for quantifying the symmetry of vector and scalar fields with respect to any irreducible representation of any symmetry group.

\section{The application of projection operators to photonic systems}\label{sec:theory}
The MS design has the symmetry of the \(C_{4v}\) point group~\cite{Willock2009,Schattschneider1978}, which gives all of the ways in which the MS can be transformed (but not deformed) such that it looks the same.  For \(C_{4v}\) this means that there are three rotation symmetries: \(\flatfrac{\pi}{2}\), known as \(C_2\); as well as both \(\flatfrac{\pi}{4}\) and \(\flatfrac{3\pi}{4}\) rotations which are both known as \(C_4\).  There are also four mirror symmetries: Horizontal and vertical, known as \(\sigma_v\); and two diagonal mirrors, known as \(\sigma_d\).  Every point group also has the identity symmetry that does nothing, known as \(E\).  Each of these symmetries can be either symmetric or anti-symmetric, but only certain combinations of symmetry and anti-symmetry are possible.  These possible combinations, known as \emph{irreducible representations}, are given in the \emph{Character table} for \(C_{4v}\) shown in \cref{tab:c4v}, where a value of 1 indicates symmetry and \(-1\) anti-symmetry and 0 that this symmetry is not present.  Note that the \(E\) irreducible representation is two-dimensional, which gives rise to its more complicated character.  All that is required to know for the purposes of this paper is that the \(E\) irreducible representation has the same symmetry as a vector.  The \(E\) irreducible representation should not be confused with the \(E\) (identity) symmetry operation.

\begin{table}[htb]
\centering
\begin{tabular}{l|rrrrr}
\hline\hline
    \(\mathbf{C_{4v}}\) & $E$ & $2C_4$ & $C_2$ & $2\sigma_v$ & $2\sigma_d$\\
     \hline
     $A_1$ & 1 & 1 & 1 & 1 & 1 \\
     $A_2$ & 1 & 1 & 1 & -1 & -1 \\
     $B_1$ & 1 & -1 & 1 & 1 & -1 \\
     $B_2$ & 1 & -1 & 1 & -1 & 1 \\
     $E$   & 2 &  0 & -2 &  0 &  0 \\
     \hline\hline
\end{tabular}
\caption{The character table for the \(C_{4v}\) point group.  Each row is a different irreducible representation, the name of which is given in the first column.}\label{tab:c4v}
\end{table}

\begin{figure}[t]
    \centering
    \vspace*{5mm}
    \includegraphics[width=0.75\columnwidth]{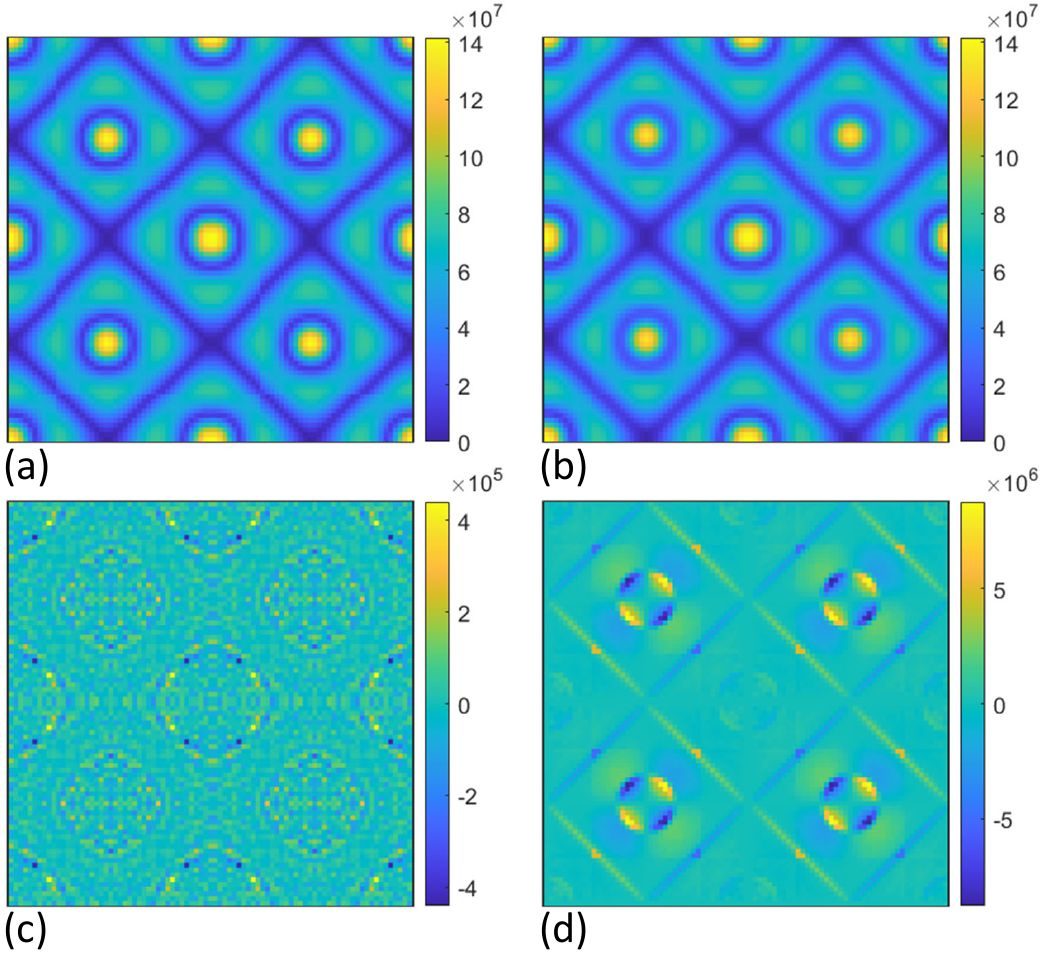}
	\caption[Effect of projection operators]{(a) \(|\vb{E}|\) (V/m) for one of the non-optimized modes in the MS, \(\ket{\psi_{A_1}}\), across four unit cells.  (b) The effect of the \(C_{4v}\ A_1\) projector on the scalar field in (a), \(\hat{P}_{4v}^{(A_1)}\ket{\psi_{A_1}}\).  (c) The effect of the \(C_{4v}\ B_1\) projector on the scalar field in (a), \(\hat{P}_{4v}^{(B_1)}\ket{\psi_{A_1}}\).  (d) The effect of the \(C_{4v}\ B_2\) projector on the scalar field in (a), \(\hat{P}_{4v}^{(B_2)}\ket{\psi_{A_1}}\).}
    \label{fig:proj}
\end{figure}

If we have a scalar field with a certain symmetry, such as that shown in \cref{fig:proj}a, then we can describe this scalar field with a vector \(\ket{\psi}\) where each element of the vector represents a different data point in the field.  That is, if we have \(n\) data points then \(\ket{\psi}\) will be an \(n\)-dimensional vector.  It is a remarkable fact that it is possible to define a hyper-plane in this \(n\)-dimensional space such that every and only vectors in this hyper-plane represent fields with a given symmetry.  This then means that we can tell if a scalar field has a given symmetry by seeing if the vector has any projection onto the corresponding hyper-plane.  The way to do this, of course, is with the linear algebra of projection operators, or projectors, which act on a vector to give the projection of that vector onto a plane.  

For example, if we wish to see if \(\ket{\psi}\) has any \(C_{4v}\ A_1\) symmetry then we multiply it by the projector for this symmetry, which we denote by \(\hat{P}_{4v}^{(A_1)}\).  If there are \(n\) data points then \(\hat{P}_{4v}^{(A_1)}\) is an \(n\times n\) matrix.  The resulting vector:
\[
\ket{\psi^\prime}=\hat{P}_{4v}^{(A_1)}\ket{\psi},
\]
must of necessity represent a field with \(C_{4v}\ A_1\) symmetry since the projection operator gives the projection onto this symmetry.

By using matrix multiplication, there is clearly a memory cost in creating an \(n\times n\) matrix for \(n\) data points, but for MSs, this is not a problem for two reasons:  Firstly, it is not necessary to take a 3D cube of data points since we can use a two-dimensional approximation with MS modes.  That is, since the lattice modes in a MS have either transverse electric or transverse magnetic fields~\cite{Sakoda:2004:OpticalProperties}, we can characterize the vector field symmetry by taking a 2D slice through the MS  (For low order modes, this approximation holds despite the asymmetry introduced by the substrate).  Secondly, a large number of data points are not required to test the symmetry of a MS -- we use \(n=43^2\) data points, but fewer would suffice.  

For cases where more data points are required, it would be possible to replace the matrix multiplication with vector multiplication, but a time cost would be incurred in doing so.  Note that this is not simply a matter of rotating or reflecting a matrix since multiple symmetry operations must be tested for in a normalized way.  Furthermore, for a non-square lattice, such as \(C_6\) symmetry, one no longer has a square array of values, and there are non-trivial operations such as a \(\flatfrac{\pi}{6}\) rotation, etc.  These resource problems are well known in imaging where projectors are used to calculate the null-space of an imaging system,~\cite{Kuo22}, and there is a sub-field of designing algorithms to efficiently create projection operators.  In our case, however, it is not the creation of the projector that is the largest cost (since this is a one-off calculation) but rather the memory-time trade-off in the application of the projector.

To give a more precise example of the application of projection operators, in \cref{fig:proj}a we can see the norm of the \(\vb{E}\) field for a non-optimized mode across four unit cells.  This scalar field has the same symmetry as the \(C_{4v}\ A_1\) irreducible representation and so we refer to it as \(\ket{\psi_{A_1}}\).  If we apply a \(C_{4v}\ A_1\) projector to this field (\(\hat{P}_{4v}^{(A_1)}\ket{\psi_{A_1}}\)) we get the result shown in \cref{fig:proj}b, which is hardly distinguishable from \cref{fig:proj}a.  We will therefore get a value close to unity (\(\eta=0.9777\)) for the symmetry parameter defined as
\begin{equation} \label{eq:eta}
\eta = \frac{\expval{\hat{P}_{4v}^{(E)}}{\psi}}{\braket{\psi}},
\end{equation}
which is a measure of what proportion of the vector projects onto the hyper-plane of \(C_{4v}\ A_1\) symmetry.  Note that we will not get \(\eta=1\) exactly due to mesh asymmetries, mode coupling, etc.

If, on the other hand, we use the projector for \(C_{4v}\ B_1\) symmetry, then we get the result in \cref{fig:proj}c, which shows \(\hat{P}_{4v}^{(B_1)}\ket{\psi_{A_1}}\).  The symmetry parameter, in this case, has a value of \(\eta=2\times10^{-6}\).   Or for \(C_{4v}\ B_2\) symmetry we get the result shown in \cref{fig:proj}d, which shows \(\hat{P}_{4v}^{(B_2)}\ket{\psi_{A_1}}\).  In this case \(\eta=7\times10^{-4}\).  It should be noted that in both \cref{fig:proj}c~and~d the resulting patterns have \(C_{4v}\ B_1\) and \(B_2\) symmetry, respectively, which must be the case since the projectors return a vector in the hyper-plane of these symmetries.

These examples demonstrate that if we apply the projector for an \emph{orthogonal} symmetry, then the symmetry parameter will be zero while applying the projector for the same symmetry will give a symmetry parameter value of 1.  Since the irreducible representations of a particular group are all orthogonal, and every uncoupled mode conforms to one of the irreducible representations of the MS point group~\cite{Sakoda:2004:OpticalProperties}, we have therefore formulated a numerical method that distinguishes between the field profiles by their symmetry.

We can use projection operators to characterize the symmetry of the field if we first select a basis for the data that is suitable for use with projection operators~\cite{McWeeny1963,Brittin1984}.  The key to this process is the design of the basis, which we refer to as the \emph{symmetrisation} of the data.

The first requirement for symmetrisation is on the distribution of the data points, which must match the symmetry to be tested.  That is, if we wish to test for \(\flatfrac{\pi}{2}\) rotation symmetry (where the rotation matrix is \(R(\flatfrac{\pi}{2})\)) then for the data point located at \(\va{q}\) there must be a corresponding data point located at \(R(\flatfrac{\pi}{2})\va{q}\) to compare it to.  We can achieve this by creating a lattice of data points from lattice vectors that are a given fraction of the MS lattice vectors.  For scalar fields, this alone is sufficient to apply projection operators (See \supbasis).

For a vector field, however, the vector components must also be symmetrised.  That is, to test for \(\flatfrac{\pi}{2}\) rotation symmetry, we must also test that the vector field at \(R(\flatfrac{\pi}{2})\va{q}\) is rotated by \(\flatfrac{\pi}{2}\) radians.  To achieve this, we define the vector field axes differently for each data point, where we denote the axes at point \(\va{q}\) by \(\hat{x}_q\), \(\hat{y}_q\) and \(\hat{z}_q\).  Taking the MS to lie in the \(x\)-\(y\) plane, the \(\hat{z}_q\) axis at \(\va{q}\) is left unchanged, but the \(\hat{x}_q\) and \(\hat{y}_q\) axes are rotated in the \(x\)-\(y\) plane such that the \(\hat{x}_q\) axis is parallel to the component of \(\va{q}\) in the \(x\)-\(y\) plane.  With this symmetrisation, the \(x\) and \(z\) components have the correct symmetry but the \(y\) components have mirror symmetries opposite to that desired.  This difference must be taken into account when applying projection operators to the \(y\) component, and as a result, \cref{eq:eta} becomes more complicated (See \supsymvec).  In the rest of the paper, we will continue to use the simpler \cref{eq:eta} for scalar fields with the understanding that the more complicated vector equation must be substituted when dealing with vector fields.

Whether or not the value for \(\eta\) is a discrete 0 or 1 or a continuous value from 0 to 1 depends both on the mode and the projector used.  The symmetry of the modes of a MS with point group \(G\) must have the symmetry of one of the irreducible representations \(g\in G\).  Therefore, if we use the projector \(\hat{P}_G^{(g)}\) we will have either
\begin{align}
    \eta =& \frac{\expval{\hat{P}_{G}^{(g)}}{\psi_g}}{\braket{\psi_g}}\\
    =& 1
    \intertext{or}
    \eta =& \frac{\expval{\hat{P}_{G}^{(g)}}{\psi_{g^\prime}}}{\braket{\psi_{g^\prime}}}\qquad g^\prime\neq g\\
    =& 0,
\end{align}
within numerical error (Including factors such as mesh asymmetry).  

There are two cases where the value of \(\eta\) will be continuous in the range \([0,1]\):  Firstly, if two modes are coupled then the symmetry of each will be a mixture of the two uncoupled symmetries, and hence one will obtain a fractional value for each of these irreducible representations.  Secondly, we have as a property of projectors~\cite{Brittin1984} that
\[
\sum_{g\in G}\hat{P}_G^{(g)}=\mathbb{1},
\]
which means that each point group spans the entire space of possible fields.  That is, any random field can be decomposed into the irreducible representations of any point group.  This then means that we are not limited to the projector that matches the point group of a MS and we can use any projector at all to quantify the symmetry of a mode.  However, if the projector has a different point group to the MS then its modes are not guaranteed to match one of the irreducible representations of the projector point group and hence we might obtain a fractional value.  

As an example of this scenario, when one moves away from the \(\Gamma\)-point, the symmetry of the MS relative to the incident field changes and so if we decompose the modes over the whole Brillouin zone with the same projector as at the \(\Gamma\)-point then one is only guaranteed to get integer values at the \(\Gamma\)-point for uncoupled modes.  The only exception being at some point away from the \(\Gamma\)-point with point group symmetry \(H\) which has that the irreducible representation \(h\in H\) is equivalent to the irreducible representation \(g\in G\) of the \(\Gamma\)-point, in the sense that
\[
\hat{P}_H^{(h)}=\hat{P}_G^{(g)}.
\]
Of necessity then, a \(h\) symmetric mode will give \(\eta=1\) for \(\hat{P}_G^{(g)}\).

Note that in the application of projectors, there is some subtlety in the choice of origin for the field (See \suporigin) as well as in the symmetry of vector fields in band-folded MSs (See \supamb).

\section{Simulations}\label{sec:sim}
To perform the optimization we used the surrogate optimization~\cite{Forrester2008} routine provided with MATLAB.  This creates a model from a number of simulations, which is then used to select the next set of parameters to simulate, which are in turn used to update the model.  All simulations were performed in COMSOL Multiphysics.

The Figures Of Merit (FOMs) for the objectives outlined in \cref{sec:design} are as follows:
\vspace{-2mm}
\begin{enumerate}[leftmargin=*]
\itemsep0em
    \item We modeled the UCNP transition as a Gaussian function at 660nm with a FWHM of 10nm so that the FOM is the value of this Gaussian function at the eigenmode wavelength.
    \item It has been shown~\cite{Sakoda2012,Mei:2012-35141:PRB} that in a \(C_{4v}\) symmetric structure Dirac cones can be formed with the overlap of \(E\) symmetric modes with either \(A_1\) and \(B_1\) or \(A_2\) and \(B_2\) symmetric modes.  These coupled modes do not rapidly drop in Q away from the \(\Gamma\)-point and are therefore excluded using our projection operator techniques.
    \doubleitem In COMSOL we integrated \(|E|^2\) over the UCNP region and divided this by the integral of \(|E|^2\) over the whole unit cell.  This was then normalized by dividing by the ratio of the volume of the UCNPs to the volume of the whole unit cell.  This normalization was included because the proportion of the unit cell taken up by the UCNP region can vary from one set of parameters to another and we wished to avoid giving a higher FOM for this reason.
    \item We modeled the absorption transition as a Gaussian function at 980nm with a FWHM of 70nm.  The MS eigenmode was modeled as a Lorentzian function at the eigenmode wavelength with the quality factor returned by simulation, scaled by the power absorption in the UCNP region for the mode (\(\omega\int nk\epsilon_0|E|^2dV\)).  Both of these functions were multiplied together and the integral of this was taken to be the FOM.
\end{enumerate}

The total FOM for the design is then the product of these five FOMs.

\section{Results}\label{sec:results}

\begin{figure}[t]
    \centering
    \vspace*{5mm}
    \includegraphics[width=0.75\columnwidth]{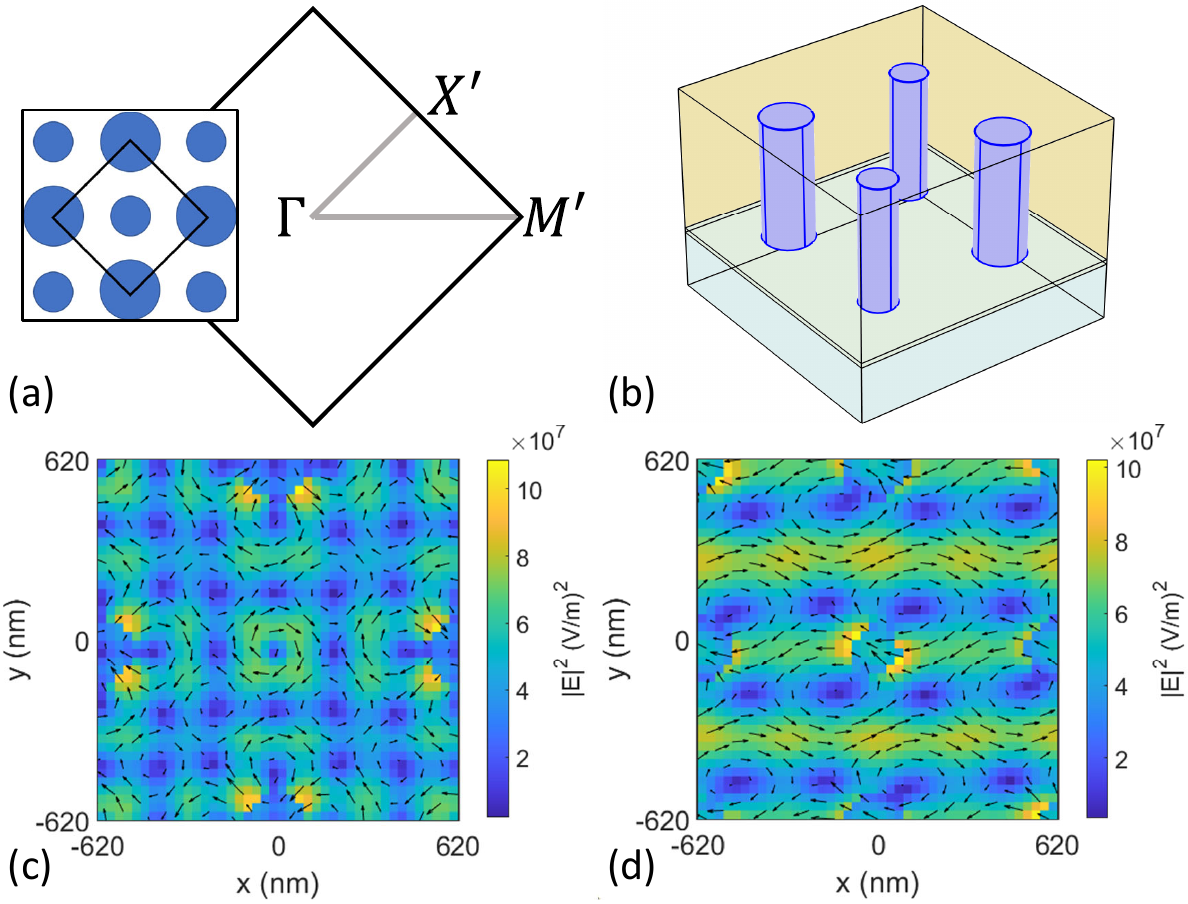}
	\caption[Results of the optimization]{(a) Nomenclature for the different directions in the Wigner-Seitz unit cell, (a, Inset) The unit cell in real space. (b) The optimized design with periodicity of 1318nm, radii of 119nm and 86nm, pillar height of 604nm and UCNP height of 704nm. (c,d) The mode selected by the optimization routine for the (c) lasing band and (d) absorption band, where the fields have been shifted to position a \(C_{4v}\) symmetry point at the center.}
    \label{fig:res}
\end{figure}

The choice of band folding for this study results in the Wigner-Seitz unit cell boundaries being rotated as well as expanded.  We therefore get that the orientation of the new \(X\) and \(M\) points in the first Brillouin zone, which we denote \(X^\prime\) and \(M^\prime\), are rotated as well.  The orientation of these points is shown in \cref{fig:res}a.  The optimized design is shown in \cref{fig:res}b where the periodicity is 1318nm, the resonator height is 604nm and that of the UCNPs is 704nm.  The two radii of the resonators are 119~nm and 86~nm.  Also shown in \cref{fig:res}b is the 23nm thick ITO layer that improves adhesion to the substrate.

At the target lasing wavelength of 660~nm the optimization routine returned a mode at 663~nm with a quality factor of 715 and the normalized ratio of the EM energy in the UCNP region to that in the whole unit cell is 5.2.  By comparison, the plasmonic MS in ref.~\cite{Fernandez-Bravo:2019-1172:NMAT} had a quality factor of 275 and a ratio of EM energy in the UCNPs of 5.9, which compares well with our result given that one would expect a surface plasmon to have little field in the resonator.  At the lasing wavelength, we get that the total electromagnetic energy in the UCNPs 
\begin{align*}
U =& \frac{1}{2}\int_{UCNP}n_{\scriptscriptstyle UCNP}^2\varepsilon_0|E|^2+\frac{1}{\mu_0}|B|^2\dd V\\
=& \int_{UCNP}n_{\scriptscriptstyle UCNP}^2\varepsilon_0|E|^2\dd V,
\end{align*}
is 6.5 times higher in our design than for that in ref.~\cite{Fernandez-Bravo:2019-1172:NMAT}. The optimization routine therefore successfully optimized for \cref{it:ldal,it:lasfield}.  \Cref{it:lasQ} was also met because the routine rejected overlapping modes of \(E\) and either \(A_1\) and \(B_1\) or \(A_2\) and \(B_2\) symmetry, which do not drop in Q away from the \(\Gamma\)-point.  We were able to do this by using the symmetry analysis technique outlined above.  A slice of the vector field taken through the center of the resonators is shown in \cref{fig:res}c.  Visual inspection shows that the field is a \(C_{4v}\ A_2/B_2\) symmetric mode, where the ambiguity is that explained in \supamb.  This is a `dark mode' that has a finite quality factor due to the substrate allowing coupling to the radiation channels as well as due to material losses.

\begin{figure}[t]
    \centering
    \vspace*{5mm}
    \includegraphics[width=0.75\columnwidth]{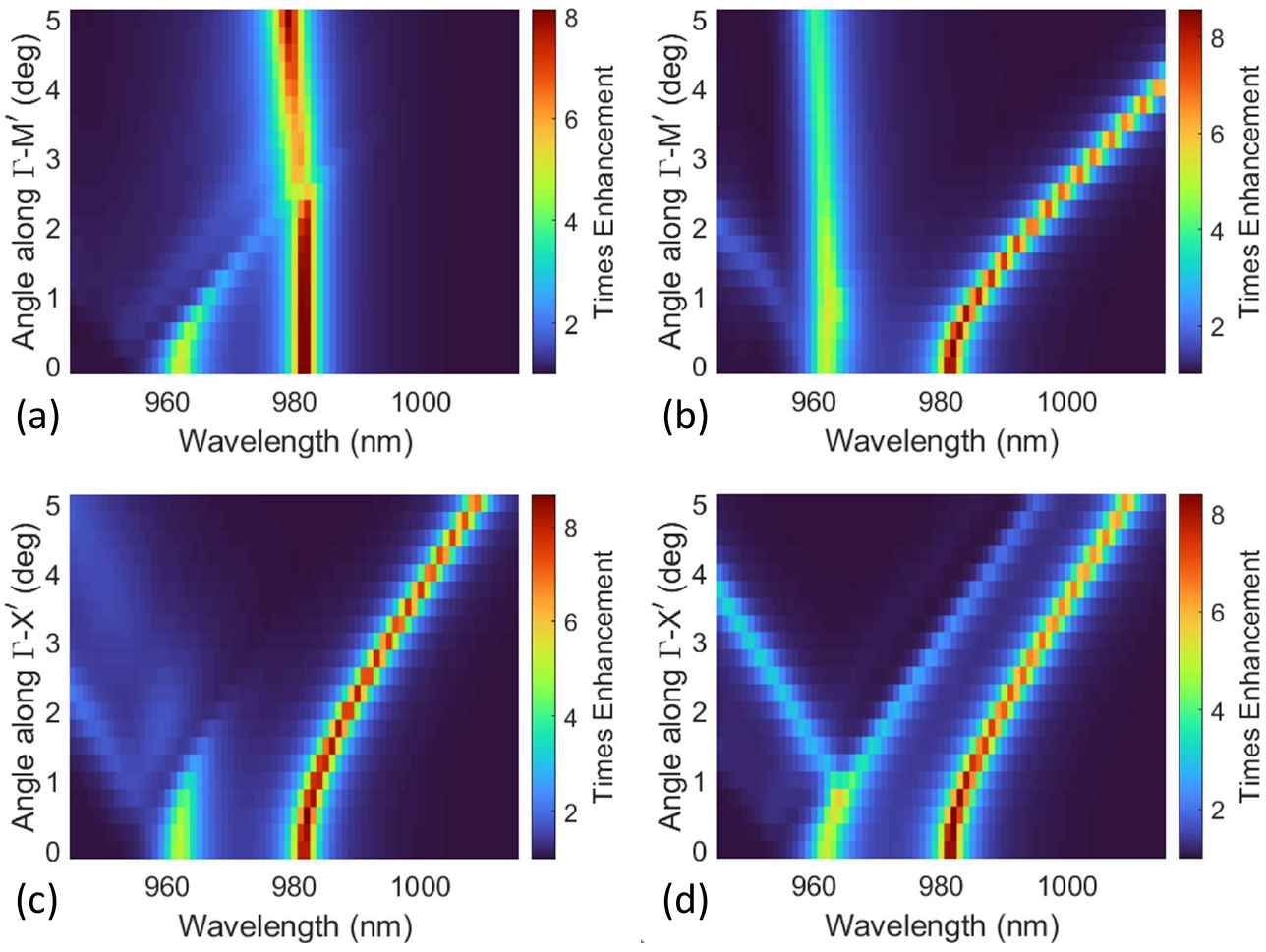}
	\caption[Results of the optimization]{The enhancement relative to a thin film of UCNPs for (a) P-polarized light along \(\Gamma-M^\prime\) (See \cref{fig:res}a), (b) S-polarized light along \(\Gamma-M^\prime\), (c) P-polarized light along \(\Gamma-X^\prime\), (d) S-polarized light along \(\Gamma-X^\prime\).}
    \label{fig:abs}
\end{figure}

At the absorption band, the selected mode is at 981.5nm with a quality factor of 300 and with a ratio of the field in the UCNPs to the whole unit cell of 6.8.  The field profile is shown in \cref{fig:res}d and clearly has \(C_{4v}\ E\) symmetry. In \cref{fig:abs} we can see the enhancement around 980nm of the power absorbed by the UCNP material relative to a film of UCNPs of the same height.  The figures show the dispersion for the \(\Gamma-M^\prime\) (a,b) and \(\Gamma-X^\prime\) directions (c,d), and for both P-polarized (a,c) and S-polarized (b,d) incident light.  By way of comparison, we find that a Fabry-Perot cavity of UCNPs between SiO\textsubscript{2} has no enhancement due to the similar refractive indices of the UCNPs and SiO\textsubscript{2}, and even with a Si substrate the enhancement would only be 2.23 times.

These figures show almost an order of magnitude enhancement of the absorption at 981~nm due to the selected mode.  Note that the enhancement of the UCNP emission would be approximately the square of this figure due to the upconversion process being sequential two-photon absorption.  

For the dispersion of the modes, we require that the mode remains within the FWHM of the UCNP transition at 980nm, which means that the permitted wavelength range for the modes is 945--1015nm (This is the wavelength range shown in \cref{fig:res}).  Furthermore, we require that the modes remain within this range for up to \(5^\circ\) tilt from the normal, since this is the acceptance angle for a typical experimental setup.  For P-polarized light along \(\Gamma-M^\prime\) there is low dispersion so that the enhancement remains very close to the target wavelength of 980nm.    
We can thus see that P-polarized light along \(\Gamma-M^\prime\) easily matches this objective due to its low dispersion.  Also, both S and P-polarized light have an acceptance angle of over \(5^\circ\) along \(\Gamma-X^\prime\).  Only S-polarized light along the \(\Gamma-M^\prime\) direction does not meet the objective, but it comes close with an acceptance angle of \(4^\circ\).   We can thus see that our requirements for \cref{it:absangle} are very close to that which was desired.

\section{Conclusion}\label{sec:conclusion}

In conclusion, we have presented a novel technique using projection operators that allows the symmetry of a scalar or vector field to be quantified.  As an example of the use of this technique, we have used it in an optimization routine to quantify the symmetry of the modes returned from eigenfrequency studies.  This allowed us to select only those modes with the symmetry that gave the response we required and we also excluded modes because the symmetry analysis showed an overlap of modes that would produce effects that we wished to avoid.  As a result, we have designed a MS with resonances at both the absorption and lasing bands of a UCNP-based MS laser.  We believe that these projection operator techniques will be of general use wherever symmetry considerations must be taken into account. 


\begin{acknowledgments}
We thank B. Cohen, E. Chan and P. J. Schuck for useful discussions on the  properties of UCNPs. This work was supported by the Australian Research Council's Centres of Excellence program (CE200100010) and the DARPA ENVision program (HR001121S0013).
\end{acknowledgments}



\nocite{McWeeny1963}
\nocite{projop}
\nocite{Ashcroft1976}
\nocite{Simon2018}

\bibliography{bibliography}


\end{document}